\documentclass[sigconf]{acmart}
\AtBeginDocument{%
  }

\copyrightyear{2026}
\acmYear{2026}
\setcopyright{cc}
\setcctype{by}
\acmConference[CIKM '26]{Proceedings of the 35th ACM International Conference on Information and Knowledge Management}{November 07--11, 2026}{Rome, Italy}
\acmBooktitle{Proceedings of the 35th ACM International Conference on Information and Knowledge Management (CIKM '26), November 07--11, 2026, Rome, Italy}
\acmDOI{10.1145/3799682.3840280}
\acmISBN{979-8-4007-2539-5/2026/11}

\begin{document}

\title{TSGuard: A Real-Time Framework for Detecting and Imputing Missing Data in Streaming Time Series}

\author{Imane Hocine}
\orcid{0000-0002-9992-577X}
\affiliation{%
  \institution{University of Luxembourg}
  \country{Luxembourg}
}
\email{imane.hocine@uni.lu}
\author{Asma Abboura}
\orcid{0000-0002-1825-0097}
\affiliation{%
  \institution{Hassiba Benbouali University of Chlef}
  \country{Algeria}
}
\email{a.abboura@univ-chlef.dz}

\author{Soror Sahri}
\orcid{0000-0002-1554-7565}
\affiliation{%
  \institution{Université Paris Cité}
  \country{France}
}
\email{soror.sahri@u-paris.fr}

\author{Abhijith Senthilkumar}
\orcid{0009-0004-5063-3882}
\affiliation{%
  \institution{University of Luxembourg}
  \country{Luxembourg}
}
\email{abhijith.senthilkumar@uni.lu}

\author{Yacine Hakimi}
\orcid{0009-0005-4820-8091}
\affiliation{%
  \institution{École Supérieure d'Informatique}
  \country{Algeria}
}
\email{y\_hakimi@esi.dz}

\author{Grégoire Danoy}
\orcid{0000-0001-9419-4210}
\affiliation{%
  \institution{University of Luxembourg}
  \country{Luxembourg}
}
\email{gregoire.danoy@uni.lu}

\renewcommand{\shortauthors}{Hocine et al.}

\begin{abstract}
Streaming sensor applications routinely suffer from delayed or missing observations caused by faults, communication losses, or environmental interference. Although recent imputation methods exploit temporal and spatial dependencies effectively, most either assume offline access to future observations or prioritize throughput without enforcing domain plausibility. We present TSGuard, a real-time demonstration system for monitoring, validating, and imputing missing values in streaming time series. TSGuard combines a lightweight graph-aware temporal imputation model with constraint-aware validation, fallback estimation, and operator-facing explanations.

Rather than treating imputation as an isolated prediction task, TSGuard integrates it into a broader data-quality loop: detect problematic observations, impute missing values, validate estimated against physical and spatial constraints, and either retain the original value as a plausible anomaly or replace it when it violates domain constraints. Using environmental sensing as a motivating setting, the demo enables users to inspect delayed sensors, compare imputers, define constraints, and validate flagged values in real time. The combination of lightweight online spatiotemporal imputation, domain-aware validation, and explicit retain-or-replace decisions is our central contribution, while interactive explanations make these decisions inspectable and actionable.
for operators.

\end{abstract}




\maketitle

\section{Introduction}
Time-series data quality problems such as incompleteness, delay, inconsistency, and invalid values are pervasive in IoT and environmental monitoring pipelines. Systems such as TsQuality \cite{tsquality} show the importance of framing time-series quality in terms of completeness, timeliness, consistency, and validity rather than as a single missing-value problem. In real deployments, however, operators need more than quality measurement: once delayed or missing observations are detected, the system must decide whether to wait, impute, raise an alert, or request human validation.

This need is particularly acute in environmental sensing, where communication failures, cloud cover, energy-saving modes, or hardware faults can interrupt data streams used for air-quality monitoring, wildfire surveillance, or hazard response. In such settings, missing values must be imputed under strict latency constraints, while remaining anomaly-aware to avoid suppressing meaningful events, preserving spatial consistency across nearby stations without propagating abnormal readings, and enforcing physically plausible values under domain constraints.

Prior work addresses parts of this problem. Classical and deep spatiotemporal imputers such as ST-MVL \cite{stmvl}, GRIN \cite{grin}, DeepMVI \cite{deepmvi}, INCREASE \cite{increase}, and PriSTI \cite{pristi} effectively model cross-sensor and temporal dependencies, but they mainly target offline or batch settings. Beyond signal-level imputation, recent work has highlighted the importance of treating time-series imputation as a broader data-quality and governance problem, where metadata, contextual relationships, and explanation mechanisms can support more reliable imputation workflows \cite{hocine2026metadata}. In contrast, online methods such as ORBITS \cite{orbits}, SANNI \cite{sanni}, and MPIN \cite{mpin} improve responsiveness for streams, but they do not combine online imputation with explicit domain constraints and operator-centric validation in a unified demo system. Meanwhile, interactive streaming tools such as I$^2$ \cite{i2} emphasize visualization rather than constraint-aware imputation. Existing approaches rarely satisfy these requirements simultaneously, especially in real-time streaming settings.

TSGuard organizes missing-value handling as a streaming quality-control loop: detect delayed or missing readings, estimate candidate values from spatial and temporal context, validate them against domain rules, compute a fallback when needed, and expose the decision to the operator.

\subsection{Motivating Example}
Consider sensors $\{S_1, S_2, S_3, S_4\}$ streaming temperature measurements every 15 minutes. At $t=3$, $S_3$ reports a sharp increase (e.g., $74^\circ\mathrm{C}$), which may correspond either to a real environmental event, such as the early stage of a wildfire, or to abnormal sensor behavior. At $t=4$, the nearby sensor $S_1$ fails to report a measurement. Standard online or batch imputation methods typically rely on recent values or learned global patterns and may therefore fail to capture rapidly evolving local conditions. A simple temporal baseline may insert a typical recent value (e.g., $22^\circ\mathrm{C}$), ignoring the abnormal rise observed at neighboring sensors, while a more reactive deep model may overestimate the missing value (e.g., $88^\circ\mathrm{C}$), exceeding realistic physical limits. This example illustrates the need for real-time imputation that is spatially consistent, anomaly-aware, and physically plausible.

\subsection{Contributions}
We summarize TSGuard’s main contributions as follows:

\begin{itemize}

\item \textbf{Streaming spatiotemporal imputation.} 
TSGuard performs real-time imputation using only past and current observations, while modeling both temporal dynamics and spatial dependencies across sensor networks, making it suitable for live environmental monitoring.

\item \textbf{Constraint- and anomaly-aware decision loop.} 
TSGuard integrates imputation with a validation pipeline that enforces physical constraints and spatial consistency, detects anomalous readings, and computes fallback estimates when candidate values violate domain rules.

\item \textbf{Operator-facing quality-control workflow.} 
Rather than treating imputation as a black-box prediction task, TSGuard exposes each decision—including validation outcomes and fallback behavior—to the operator, enabling inspection, explanation, and human-in-the-loop validation.

\item \textbf{End-to-end interactive demo system.} 
We present TSGuard as a fully integrated system that combines detection, imputation, validation, visualization, and user interaction, supporting real-time monitoring and exploration of environmental sensor data.

\end{itemize}
TSGuard has been validated on Beijing AQI-36 data \cite{stmvl} and is publicly available at \href{https://github.com/kaaaziz/ts_guard}{the project repository} 
A demonstration video is also available \href{https://youtu.be/WPb3QIl_MRU}{here}.

\section{TSGuard Architecture}
\begin{figure}[b]
    \centering
    \includegraphics[width=0.98\columnwidth]{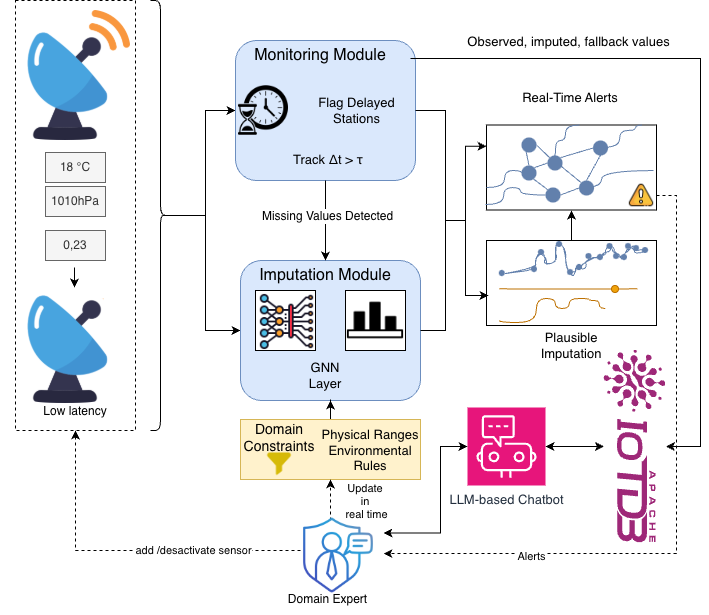}
    \caption{TSGuard architecture: monitoring, streaming imputation, constraint validation, and operator-facing explanation are integrated in one loop.}
    \label{fig:arch}
\end{figure}

Figure~\ref{fig:arch} summarizes TSGuard. The architecture mirrors the contribution structure above, so that the system design stays aligned with the paper message.

\subsection{Monitoring and Quality Assessment}
The monitoring module continuously ingests sensor readings, aligns them by timestamp, and tracks timeliness. Let $t_{i,\mathrm{last}}$ denote the most recent arrival time for sensor $S_i$, and let $\Delta t_i$ denote its current delay. A sensor is flagged as delayed when $\Delta t_i > k$, where $k$ is an application-defined threshold. This framing keeps the discussion general and consistent with time-series quality notions such as completeness, timeliness, and validity \cite{tsquality}.

TSGuard exposes two classes of domain constraints. \emph{Physical-range constraints} encode acceptable value intervals (for example, plausible temperature or AQI ranges). \emph{Spatial-consistency constraints} limit acceptable deviations between neighboring sensors. These rules are applied to both incoming readings and imputed values. The goal is not to replace the imputer with rules, but to make the decision process auditable and robust.

\begin{figure*}[hbt]
    \centering
    \setlength{\tabcolsep}{3pt}

    \resizebox{0.7\textwidth}{!}{%
    \begin{tabular}{@{}p{0.25\textwidth} p{0.25\textwidth} p{0.25\textwidth} p{0.25\textwidth}@{}}
        
        \multicolumn{3}{p{0.75\textwidth}}{%
            \fbox{\includegraphics[width=0.99\linewidth]{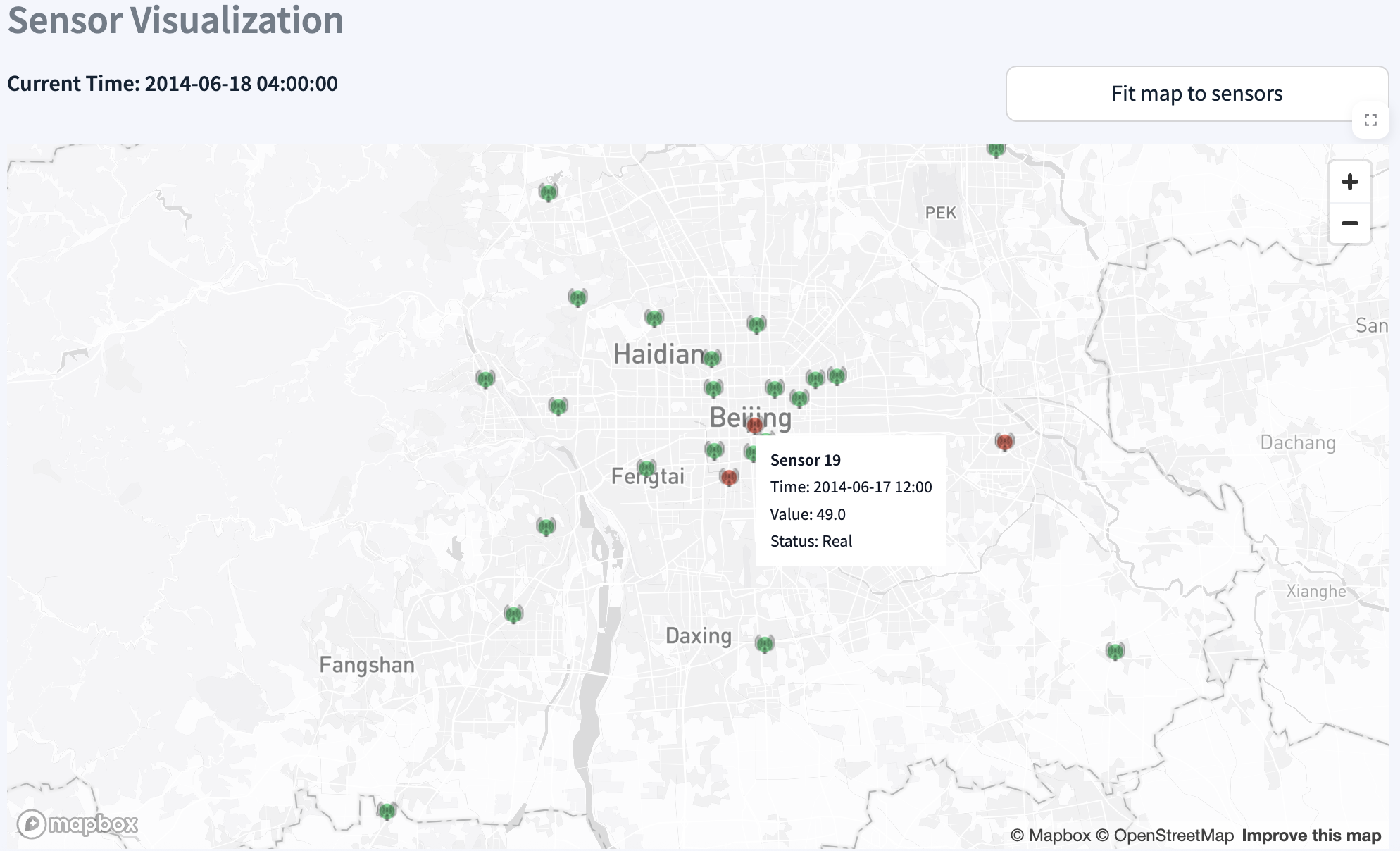}}
        }
        &
        \begin{tabular}{@{}p{\linewidth}@{}}
            \vspace{-7.95cm}
            \fbox{\includegraphics[width=\linewidth]{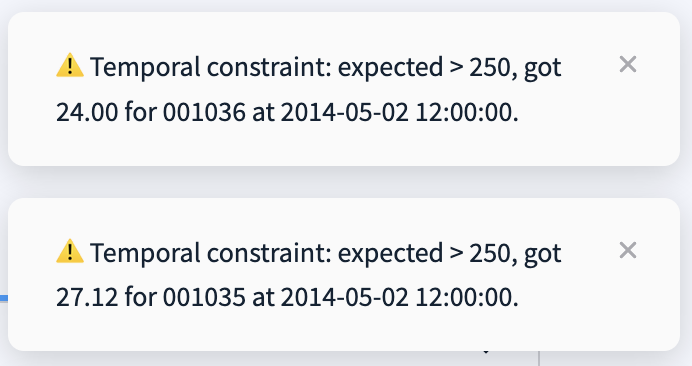}} \\

            \centering
            \vspace{-5.7cm}
            \fbox{\includegraphics[width=\linewidth]{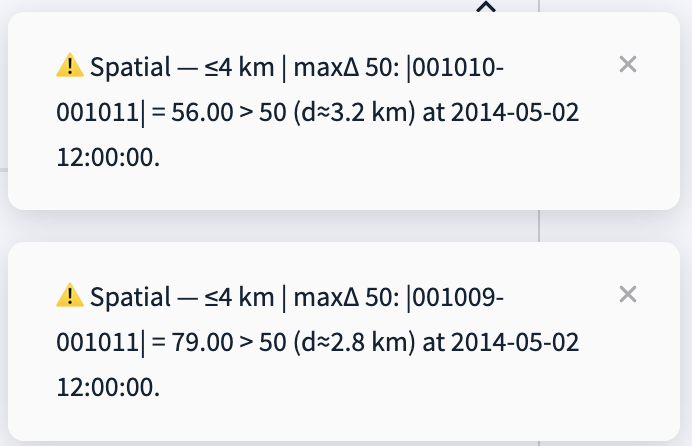}} \\
            \fbox{\includegraphics[height=2.2cm,width=\linewidth]{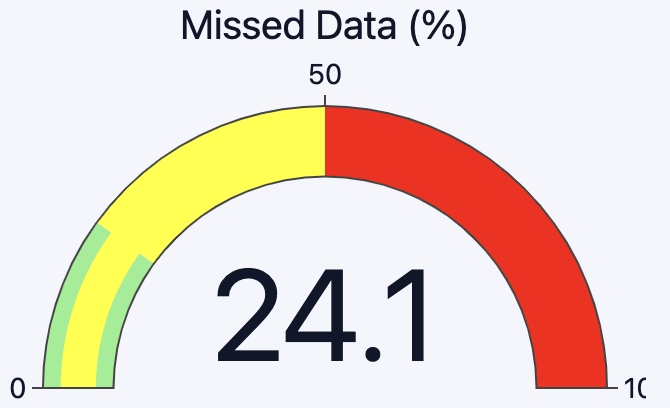}}\\
        \end{tabular}
        \\

        \multicolumn{2}{p{0.50\textwidth}}{%
            \fbox{\includegraphics[height=4.22cm,width=\linewidth]{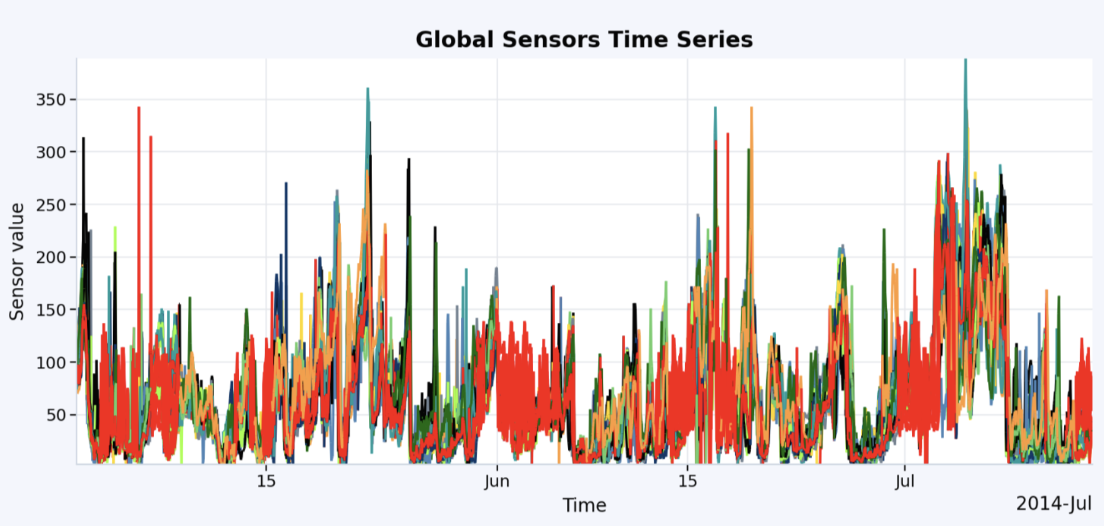}}
        }
        &
        \multicolumn{2}{p{0.50\textwidth}}{%
            \fbox{\includegraphics[height=4.22cm,width=\linewidth]{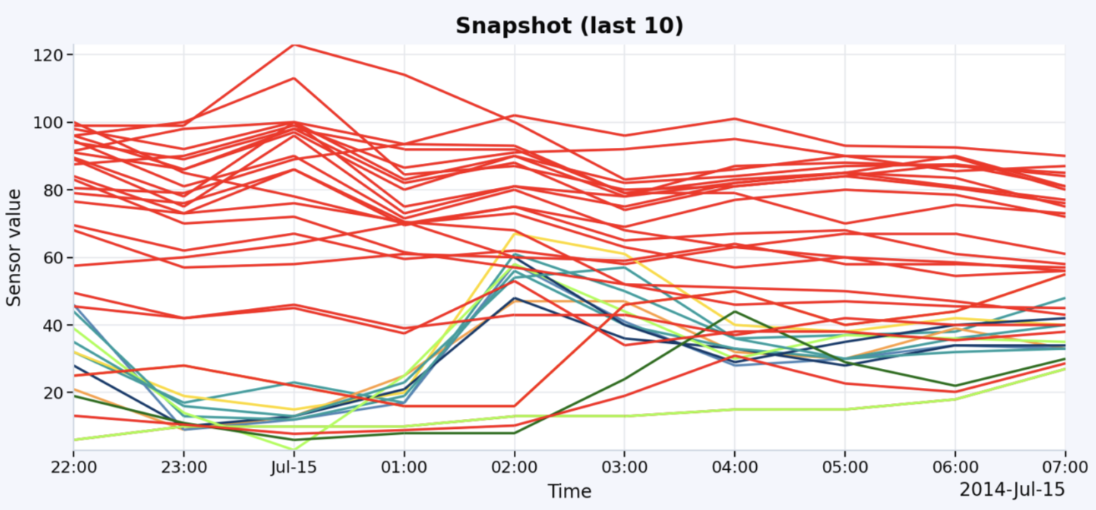}}
        }
    \end{tabular}%
    }

    \caption{Selected screenshots of TSGuard.}
    \label{maininterface}
\end{figure*}

\subsection{Streaming Imputation Engine}
Once a missing or delayed reading is confirmed, TSGuard invokes a hybrid graph-temporal imputer. A graph neural layer aggregates neighborhood context so that nearby or correlated sensors can influence imputation; the resulting embeddings are then processed by an LSTM that captures short- and medium-term temporal dynamics. A final prediction layer emits a candidate value for the missing reading.

This choice intentionally differs from the positioning of PriSTI \cite{pristi}. PriSTI uses a conditional diffusion design that achieves strong offline accuracy by leveraging rich context. TSGuard instead favors a lighter pipeline that can run incrementally, without future observations, and can therefore respond within interactive latency. Relative to GRIN \cite{grin} and INCREASE \cite{increase}, TSGuard also prioritizes the deployment setting: the model is embedded in a continuous monitoring pipeline rather than being used as a standalone research predictor.

\subsection{Constraint Validation and Fallback}
Candidate imputations are validated immediately after prediction. If a value violates user-defined constraints, TSGuard computes a fallback estimate from valid neighbors. Let $N_i$ be the neighbor set of sensor $S_i$ and let
\[
V_i(t)=\{S_j\in N_i \mid v_j(t)\;\text{satisfies constraints}\}.
\]
TSGuard computes
\[
\hat v_i^{\mathrm{fb}}(t)=\frac{\sum_{S_j\in V_i(t)} w_{ij}v_j(t)}{\sum_{S_j\in V_i(t)} w_{ij}},
\]
where $w_{ij}$ captures spatial proximity and recent temporal similarity. If no valid neighbor is available, the system reuses the last valid value of $S_i$ as a temporal fallback.

This validation layer is a key component of TSGuard, because it connects imputation with alerting, fallback computation, and interactive decision support.

\subsection{Alerting and Explanation}
All delayed sensors, constraint violations, and implausible imputations are forwarded to the alert module. The interface explains why a value was flagged, shows the fallback estimate when one exists, and lets the operator keep the original value as a real anomaly or replace it with the fallback. This preserves potentially important extreme events instead of silently smoothing them away, a point that is especially relevant for environmental hazards and anomaly-aware monitoring \cite{tsad_overview}. 

To support traceability and operator interaction, 
all incoming, validated, and imputed time-series values are durably persisted in Apache IoTDB \cite{tsquality}, enabling low-latency access for real-time visualization, alert management, and retrospective analysis.


        





\section{Demonstration Overview}
\subsection{Interface and User Workflow}
Figure~\ref{maininterface} shows the demo interface. Users can upload a dataset, configure constraints, start or stop sensors, inspect alerts, and compare observed and imputed values over time. The map view exposes the spatial context, while the time-series panels show global evolution and recent snapshots.

The demonstration follows four scenarios.
\textbf{(1) Delay below or above threshold.} A delayed sensor is first monitored without imputation; once the delay exceeds $k$, TSGuard imputes the missing reading and validates it. \textbf{(2) Missing value with unreliable neighbors.} If nearby sensors are also unavailable, TSGuard falls back to temporal history and raises a stronger alert. \textbf{(3) Dynamic sensor management.} Users can deactivate or add sensors at runtime to emulate maintenance or coverage expansion. \textbf{(4) Audience-defined constraints.} Attendees can change domain rules and immediately observe how they affect alerting and imputation.

The demo emphasizes a system-level message that is often implicit in algorithm papers: in streaming environments, the practical question is not only which model has the lowest error, but also whether its outputs are timely, plausible, inspectable, and actionable. TSGuard makes these trade-offs visible to attendees.

\subsection{Interactivity}
Attendees can inspect individual sensors, switch between observed and imputed values, and replay periods with higher missingness to understand how the system reacts under stress. When TSGuard raises a plausibility alert, the operator can compare the model prediction with the fallback estimate and decide whether the event should be preserved as a genuine anomaly. This interaction is important for hazardous situations, where blindly correcting all extreme values would hide potentially meaningful events.

The demo also exposes the difference between \emph{quality measurement} and \emph{quality repair}. A user can tighten or relax range and spatial-consistency thresholds and immediately observe how this changes the number of alerts, the fallback behavior, and the visual state of the dashboard. In that sense, TSGuard is meant to be both a research demo and a discussion tool for data-quality policies in streaming time-series systems.

The interface also includes an LLM-based assistant (served via the Groq API for low-latency inference) that provides natural-language explanations of system behavior. The assistant is grounded in TSGuard’s internal state, including recent sensor readings, detected delays, constraint violations, and fallback decisions. Users can query the system (e.g., “Why was sensor $S_1$ flagged?”), and the assistant generates explanations that reference constraint violations, neighboring sensors, and fallback computations. This assistant has no role in the imputation or keep-or-replace decisions, which are determined independently by TSGuard’s decision logic. Thus, an incorrect LLM response cannot alter the system’s decisions or stored data.

\subsection{Evaluation Summary}

We evaluate TSGuard on Beijing AQI-36 using the same data splits and missing patterns as PriSTI \cite{pristi}, processed sequentially to emulate streaming conditions. Table~\ref{tab:eval} compares TSGuard with PriSTI, a past-only online adaptation of PriSTI (PriSTI-ON), and ORBITS. PriSTI serves as an offline accuracy upper bound, whereas PriSTI-ON and ORBITS are online-capable baselines.

PriSTI-ON is not a retrained model; it uses the released PriSTI checkpoint \cite{pristi} with the authors' original configuration and hyperparameters unchanged. To evaluate its behavior under streaming constraints, we mask all future timestamps beyond the current time step t (i.e., the current time step) in the input matrix, so that only information available up to t is provided to the model. No other model components or hyperparameters are modified. PriSTI-ON is evaluated in the same experimental environment and under the same evaluation protocol as TSGuard. The resulting performance therefore quantifies the impact of imposing a causal masking constraint on an offline diffusion imputer, rather than the performance of a diffusion model specifically trained for streaming inference. We report PriSTI-ON as an untuned causal reference baseline; designing and retraining a diffusion model specifically for streaming inference is outside the scope of this demo.

\begin{table}[t]
\centering
\small
\caption{Accuracy, relative speed, and latency (16-core CPU, batch=1).}
\label{tab:eval}
\begin{tabular}{lcccc}
\toprule
Method & MAE & RMSE & Relative Speed & Latency (ms) \\
\midrule
PriSTI \cite{pristi} & 9.03 & 17.62 & - & - \\
PriSTI-ON & 75.23 & 111.67 & 0.01 & 6000 \\
ORBITS \cite{orbits} & 18.16 & 29.35 & 1.2 & 50 \\
TSGuard & 16.13 & 28.37 & 1 & 60 \\
\bottomrule
\end{tabular}
\end{table}

TSGuard achieves competitive imputation quality among online-capable methods while maintaining interactive latency. It processes each missing observation in $\approx 60$ ms, with a throughput of $\sim 180$ sensor updates/sec on a 16-core CPU, sufficient for real-time monitoring. ORBITS is slightly faster (50 ms, $\sim 220$ sensors/sec) but lacks constraint-aware validation and fallback, leading to less robust imputations. PriSTI-ON is considerably slower ($\approx 6000$ ms, $<1$ sensor/sec) and suited for past-only adaptation, while PriSTI serves only as an offline accuracy reference.

From a demo perspective, PriSTI provides an offline upper bound, ORBITS a lightweight online baseline, and TSGuard demonstrates the benefits of combining spatiotemporal imputation with constraint enforcement and operator feedback. Although evaluated on air quality data, TSGuard applies broadly to streaming sensor domains such as industrial and IoT monitoring.

We also evaluated simplified TSGuard variants without constraint validation or fallback. Both components improve robustness by reducing implausible imputations and stabilizing predictions under missing or noisy neighborhood conditions. Detailed results, including per-module latency and throughput, are in the artifact repository.

\section{Conclusion}
TSGuard demonstrates that missing-value handling in streaming time series should be presented as a closed-loop data-quality task rather than a standalone prediction benchmark. The system combines online graph-temporal imputation, constraint-aware validation, fallback estimation, and human-in-the-loop inspection in a coherent architecture. This positioning also clarifies the contribution relative to prior work: TSGuard complements quality-oriented demos such as TsQuality with actionable imputation, and complements imputation algorithms such as PriSTI and ORBITS with real-time, constraint-aware operator support.

Future work will include adaptive constraints, tighter anomaly/ imputation integration and operator-guided parameter tuning. We will also investigate broader evaluations across additional and larger datasets and sensing domains to assess generality, latency, and throughput trade-offs.



\begin{acks}
This work is funded by the Luxembourg National Research Fund (FNR), INTER programme under the UltraBO Project (ref. INTER/ ANR/22/17133848), and the Polish National Centre for Research and Development (NCBR) (ref. POLLUX-XI/15/Serenity/2023). 
\end{acks}

\section*{Declaration on Generative AI}
The authors declare that generative AI tools (ChatGPT) were used for language refinement, including Grammar and spelling check and improve writing style. All content was subsequently reviewed and edited by the authors, who take full responsibility for the accuracy, originality, and claims presented in this work.

\bibliographystyle{ACM-Reference-Format}
\bibliography{sample-base}

@String{Springer = "Springer-Verlag" }

@article{deepmvi,
  title={Missing value imputation on multidimensional time series},
  author={Bansal, Parikshit and Deshpande, Prathamesh and Sarawagi, Sunita},
  journal={arXiv preprint arXiv:2103.01600},
  year={2021}
}

@article{grin,
  title={Filling the g\_ap\_s: Multivariate time series imputation by graph neural networks},
  author={Cini, Andrea and Marisca, Ivan and Alippi, Cesare},
  journal={arXiv preprint arXiv:2108.00298},
  year={2021}
}

@inproceedings{i2,
  title={{I}$^2$: Interactive Real-Time Visualization for Streaming Data},
  author={Traub, Jonas and Steenbergen, Nikolaas and Grulich, Philipp M. and Rabl, Tilmann and Markl, Volker},
  booktitle={Proceedings of the 20th International Conference on Extending Database Technology ({EDBT})},
  pages={526--529},
  year={2017}
}

@article{orbits,
  title={ORBITS},
  author={Khayati, Mourad and Arous, Ines and Tymchenko, Zakhar and Cudr{\'e}-Mauroux, Philippe},
  journal={Proceedings of the VLDB Endowment},
  volume={14},
  pages={294--306},
  year={2020}
}

@article{mpin,
  title={Missing value imputation for multi-attribute sensor data streams via message propagation},
  author={Li, Xiao and Li, Huan and Lu, Hua and Jensen, Christian S and Pandey, Varun and Markl, Volker},
  journal={Proceedings of the VLDB Endowment},
  volume={17},
  number={3},
  pages={345--358},
  year={2023},
  publisher={VLDB Endowment}
}

@inproceedings{pristi,
  title={Pristi: A conditional diffusion framework for spatiotemporal imputation},
  author={Liu, Mingzhe and Huang, Han and Feng, Hao and Sun, Leilei and Du, Bowen and Fu, Yanjie},
  booktitle={2023 IEEE 39th international conference on data engineering (ICDE)},
  pages={1927--1939},
  year={2023},
  organization={IEEE}
}

@article{tsad_overview,
  title={Time-series anomaly detection: Overview and new trends},
  author={Liu, Qinghua and Boniol, Paul and Palpanas, Themis and Paparrizos, John},
  journal={Proceedings of the VLDB Endowment (PVLDB)},
  volume={17},
  number={12},
  pages={4229--4232},
  year={2024}
}

@inproceedings{stmvl,
  title={ST-MVL: filling missing values in geo-sensory time series data},
  author={Yi, Xiuwen and Zheng, Yu and Zhang, Junbo and Li, Tianrui},
  booktitle={Proceedings of the 25th international joint conference on artificial intelligence ({IJCAI}},
  year={2016}
}

@inproceedings{increase,
  title={Increase: Inductive graph representation learning for spatio-temporal kriging},
  author={Zheng, Chuanpan and Fan, Xiaoliang and Wang, Cheng and Qi, Jianzhong and Chen, Chaochao and Chen, Longbiao},
  booktitle={Proceedings of the ACM Web Conference 2023},
  pages={673--683},
  year={2023}
}

@article{sanni,
  title={SANNI: Online Imputation of Missing Values in Multivariate Time Series Based on Deep Learning and Behavioral Patterns},
  author={Yurtin, AA and Zymbler, ML},
  journal={Lobachevskii Journal of Mathematics},
  volume={45},
  number={11},
  pages={5948--5966},
  year={2024},
  publisher={Springer}
}

@article{tsquality,
  title={TSQuality: measuring time series data quality in Apache IoTDB},
  author={Qiu, Yuanhui and Fang, Chenguang and Song, Shaoxu and Huang, Xiangdong and Wang, Chen and Wang, Jianmin},
  journal={Proceedings of the VLDB Endowment},
  volume={16},
  number={12},
  pages={3982--3985},
  year={2023},
  publisher={VLDB Endowment}
}

@inproceedings{hocine2026metadata,
  title={Metadata-Guided Diffusion and {LLM}-Orchestrated Quality Governance for Time Series Imputation},
  author={Hocine, Imane and Abboura, Asma and Kella, Abdelaziz and Hanini, Maria and Danoy, Gr{\'e}goire},
  booktitle={Proceedings of the 1st International Workshop on Quality in Large Language Models and Knowledge Graphs ({QuaLLM-KG}), co-located with the EDBT/ICDT 2026 Joint Conference},
  year={2026},
  publisher={CEUR-WS.org}
}
\end{document}